# BIO-THENTIC CARD: AUTHENTICATION CONCEPT FOR RFID CARD

*Ikuesan Richard Adeyemi*
*Dept. computer science and information system*
Universiti Teknologi, Malaysia
Johor Bahru, Malaysia
Raikuesan2@livew.utm.my

*Norafida Bt, Ithnin*
*Dept. computer science and information system*
Universiti Teknologi, Malaysia
Johor Bahru, Malaysia
afida@utm.my

**Abstract**
Radio frequency identification (RFID) is a technology that employs basic identifier of an object embedded in a chip, transmitted via radio wave, for identification. An RFID Card responds to query/interrogation irrespective of '*Who*' holds the Card; like a key to a door. Since an attacker can possess the card, access to such object can therefore be easily compromised. This security breach is classified as an unauthorized use of Card, and it forms the bedrock for RFID Card compromise especially in access control. As an on-card authentication mechanism, this research proposed a concept termed Bio-Thentic Card, which can be adopted to prevent this single point of failure of RFID Card. The Bio-Thentic Card was fabricated, tested and assessed in line with the known threats, and attacks; and it was observed to proffer substantive solution to unauthorized use of RFID Card vulnerability.

**Key words**: Vulnerability, unauthorized, mitigation, authentication, communication, access control system

## I. INTRODUCTION

Radio frequency identification (RFID) technology is a technology that has gained wider adoption into the human everyday life since its first usage in identification friend or foe (IFF) during the II world war [1, 3]. RFID is characterized by its ubiquitous nature, flexibility, mobility and integratability, which has contributed to its adoption in places such as access control system, conveyor control system, banking notes, item identification e.t.c. While RFID pros have greatly improve other technology, its cons has also generated series of security and privacy challenges [2, 3, 6] some to the detriment of the system being integrated into [4, 5]. However, such challenges are not limited to only RFID systems, but peculiar to RFID systems, are attacks such as relay attack, cloning, clandestine tracking, unauthorized tag read, and unauthorized tag use [2, 6, 8]. Un-authorization of card use is a general challenge in access control system; hence, most systems would require a secondary control mechanism.

However, the integration of the RFID tag into access control Card otherwise known as RFID Card has further complicated the challenges in access control cards leading to greater trade-offs in security and privacy [3, 6, 7, 8]. Access control RFID card do not provide on-card authentication system hence is openly vulnerable to attacks that breach the confidentiality of a secured system. RFID Card responds to interrogation from an RFID Reader irrespective of '*who*' holds the card, or whether the subject has the required privilege to do so. This lack of authorization priori to interrogation can be said to be the principal point of failure of the RFID Card. For instance, consider the situation where an unauthorized subject with malicious intent or the otherwise, gains access to a classified data through a stolen RFID Card and consequently jeopardize the confidentiality of the system under protection. It suffixes to note that, to the best of our knowledge, no known countermeasure addressed this single point of failure of the RFID Card.

However, mitigating this critical point of failure is not as trivial as it sounds. Faraday shield model in [1] is popular method (aluminum-foiled wallet for example) of shielding the RFID Card from unauthorized tag reading, thus enhancing the privacy protection of the RFID tag. The unauthorized tag use as applied to RFID Card is the main goal of this paper as analyzed in [31]. The remaining of this paper is organized as follows. Section II highlights the related research works on RFID tag with reference to its physical layer, discusses the principal point of failure of the RFID Card. Section III introduces the concept used in this study, detailed the design and result of this study. Section IV presents the analysis and the conclusion of this concept.

## II. RELATED RESEARCH

RFID Card is a composition of antenna unit, memory unit, processing unit, and a tag, which communicates with an RFID Reader wirelessly using the near field coupling principle. Over the past decades researchers have worked extensively on the RFID system but interest on RFID on-card authentication system have received minimal attention. According to [4, 9], the physical layer of the RFID system is the perimeter defense line for security tightening in RFID system. RFID authentication protocols [10, 11, 12, 13, 14] are designed to mitigate communication attacks between the tag and the Reader. Similarly, various lightweight cryptographic protocols and techniques [15, 16, 17, 18, 19, 20,] have also been designed to combat security vulnerabilities in the RFID system. However, these authentication practices do not apply to the tag end of the physical layer of the RFID system. Additionally, techniques such as blocker tag [21], RFID guardian [22], RFID zapper [23], Faraday shield [30] and clipped tag [24] are mitigation to distance attacks, which does not necessarily translate affect RFID cards due to short range of communication. However, in [25], a framework for user's authentication procedure was modeled using fingerprint authentication through reader-system authentication process, a similar process to [26] which is adopts a two-factor authentication system based on combined fingerprint recognition and smart RF Card verification. They however failed to address the underlying problem of the on-Card authentication of the RFID card. In [27, 28] different categories of RFID card suitable for different security integration were designed but they lacked the core and essential component of card security: user authentication. [31] gives a detailed analysis of the challenges in RFID card with reference to its physical layer. Table 1 gives the summary of the various countermeasures proposed against the physical layer authentication vulnerabilities.

Table 1: Countermeasure to physical layer Authentication Challenges

| Proposed Counter-Measure | Authentication at Physical-Layer Vulnerabilities | | | | | | | | |
|---|---|---|---|---|---|---|---|---|---|
| | Unauthorized Card reading | Unauthorized Card | Tag cloning | Relay attack | Skimming | Spoofing | Unauthorized killing | Clandestine tracking | Physical layer Identification |
| Physical-Layer Identification technique | × | × | √ | × | × | × | × | × | √ |
| Faraday Cage | √ | × | × | × | √ | √ | √ | √ | × |
| Authentication protocol | √ | × | × | √ | √ | √ | √ | √ | × |
| Clipped Tag | √ | × | × | √ | √ | √ | × | √ | × |
| Anti-counterfeiting | √ | × | √ | √ | √ | √ | × | × | × |
| Biometric authentication | × | × | × | × | × | × | × | × | √ |
| Labeling | ONLY create awareness for users | | | | | | | | |
| Controllable Tag | √ | × | × | √ | √ | √ | √ | √ | × |

The physical layer identification technique [29] addresses cloning of tags, and proves that no two tags can have the same fingerprint. Furthermore, controllable tag[27] addressed the issue of unauthorized tag read, thus curbing one of the principal source of attacks on RFID tag. However, on-card authentication vulnerability, which is a major security challenge, have received little or no attention as shown in Table 1 Countermeasure such as clipped tag, and fingerprint biometric authentication [25] can be combined in a digitalized manner to curtail this challenge. In the next session we, present our concept of Bio-Thentic Card as a concept of On-Card authentication process, which is a combination of digitalized controllable clip tag and fingerprint authentication system.

### III. ON-CARD AUTHENTICATION CONCEPT

The architecture of the RFID Card reveals that communication between the Card and the Reader is hinged on the interconnection between the antenna unit and the tag *inside of the* Card. The antenna (usually rectangular spiral) unit of the RFID card is the medium of interaction between the tag of the RFID Card and the RFID Reader. Hence, the connectivity, transmission range and power supply to the RFID tag is a function of the antenna unit. Suppose we represent the communication process as $C_p$ which is the integration of the antenna unit joints $(A_{uj})$, and the RFID tag $(R_t)$. For the sake of this paper, we represent every other parameter surrounding the RFID tag such as battery, memory unit, as RFID tag. We also assume that the antenna unit is the suitable antenna for RFID card. The communication process, $C_p$ is given by *equation (i)*.

$$C_p = \sum_{i=0}^{k} ( \sum_{j=0}^{n} R_t \text{ x } A_{uj} ) \quad (1)$$

If $A_{uj} = 0$, then, the communication process $C_p$ presented in equation *(i)* becomes:

$$C_p = \sum_{i=0}^{k} ( \sum_{j=0}^{n} R_t \text{ x } 0 ) = 0 \quad (2)$$

This illustrates that if the possible contact between the RFID tag and the antenna unit can be disconnected such that the total corresponding antenna unit connection is zero, then, the antenna communication process $(C_p)$ will be zero. With this criteria, we observed that the unauthorized use of card vulnerability in the RFID Card can be mitigated using the combination of digitally clippable tag-antenna-joint, and a biometric authentication system, preferably, fingerprint, as analyzed in [31]. Furthermore, we observed that a strategic placement of a digitally controllable hinge between the antenna and the tag in such as way that the antenna forms a shield around the tag, when totally disconnected from the tag, will prevent privacy disclosure, tracking and all radio wave related attacks. When this clippable joint is then strictly controlled by an authentic subject, the single point of failure of the RFID Card can thus be mitigated. We termed this concept Bio-Thentic Card (BTC), which is the integration of biometric component into the RFID tag

## IV. RESEARCH METHOD

Our research aimed at conceptualizing an RFID Card (which we called BTC) which can mitigate the unauthorized Card use vulnerability. In order to achieve our aim, we designed our methodology into three distinct stages.

*Stage1*: this stage comprises the design, calibration, simulation and fabrication of the card antenna unit. In this stage, we analyzed thoroughly; the suitable positioning, and control of the clippable joint, such that the Card will respond to interrogation only through the contact from the clip joint.

*Stage2*: this stage involves the acquiring, authenticating, securing and storage of the biometric authentication process, fingerprint in this case. We carefully considered the choice of the fingerprint module to use in line with information security practices such as security of the fingerprint module (live fingerprint detection, and false error rate) and secure code development practice.

*Stage3*: this stage involves the integration of the various modules, and the control module. The result and testing process is detailed in the next session. The control unit integrates the biometric fingerprint and the fabricated antenna unit into a single module controlled by a microcontroller. Figure 1 gives a detailed description of the our designed methodology

The Output from the clip joint and the Faraday cage must be 'Yes' before stage1 can be passed to stage2 as shown in Figure 1. The communication between stages 1, 2 and 3 is illustrated in Figure 2. We designed a rectangular loop antenna consisting of stripped copper lines, with external dimension of 54x33mm, 0.5mm width, 7 turns, 1mm spacing and 0.035mm thickness using a computer simulation technology (CST) studio as shown in Figure 3 and 4. The design comprises a PCB made of FR4-lossy dielectric material with thickness of 1.6mm, dimension of 60x40mm, relative permeability of 1, and relative electric permittivity of 4.55.

We integrated the clipped joint as shown in Figure 3 through the fabrication process of the card antenna unit with a 13.56MHz RFID tag (see Figure 5). The digitalized controllable hinge was introduced through a miniature relay of 1A, 5V direct current, and internal coil resistance of 166ohms. Upon simulation, we arrived at an S-parameter value of -2.730712, which we considered as suitable for our experimental purpose as illustrated in Figure 4.

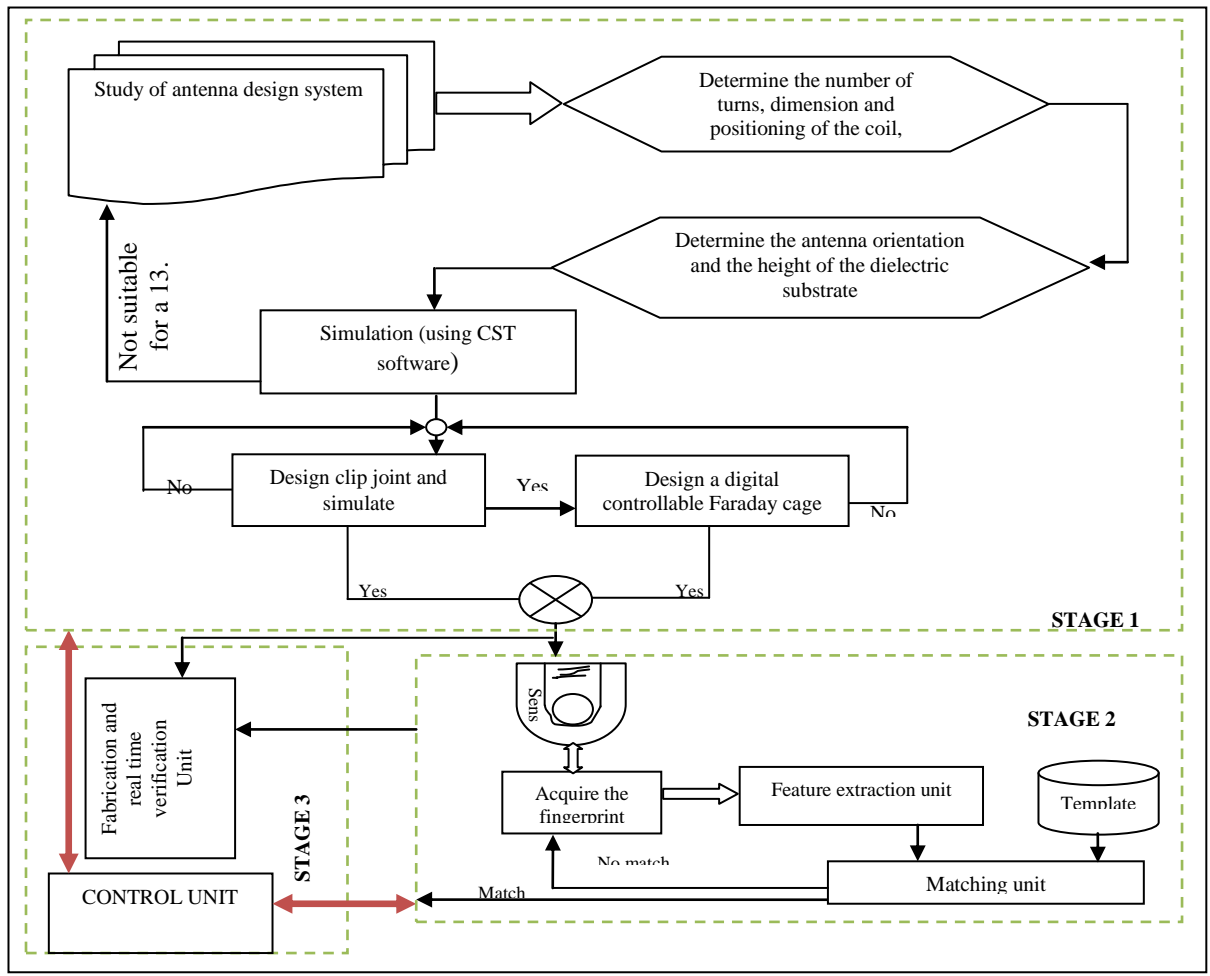

Figure 1: Design Flow Process

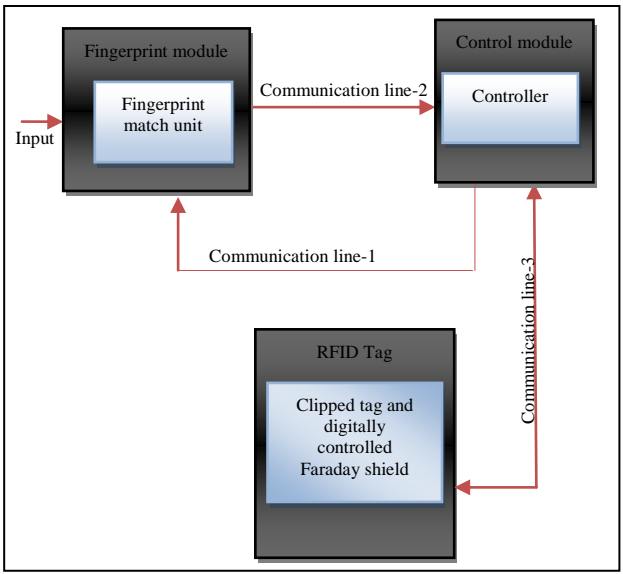

Figure 2: Communication process of the Bio-Thentic Card

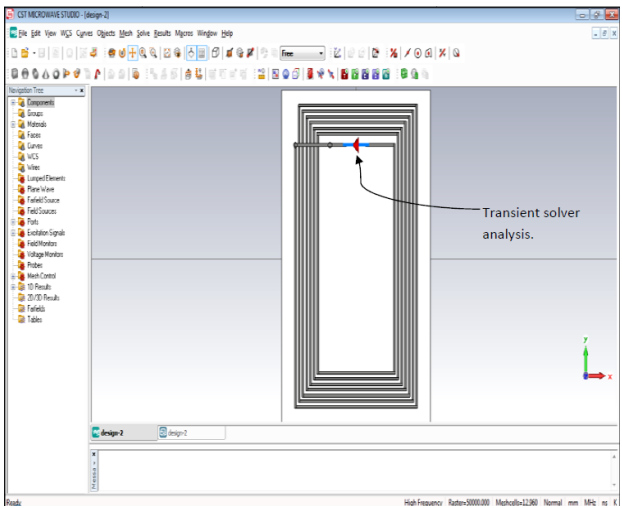

Figure 3: Antenna structure

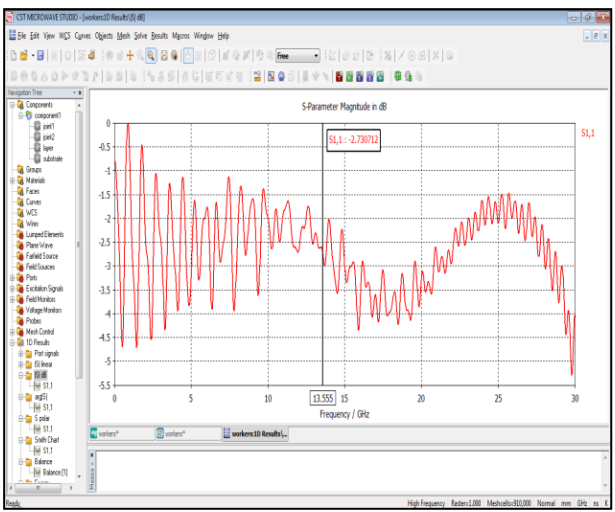

Figure 4: S-parameters as a function of frequency

## V. BIO-THENTIC CARD (BTC)

A secured fingerprint module was adopted for the biometric authentication process. Moreover, it was designed as an on-card biometric match system. Two distinct fingerprints of the authentic user are required for the operation of the Card. Additionally; we stored other fingerprints templates for testing purpose, and tagged them with various identities. The communication process shown in Figure 4 depicts the link between the fingerprint module, and the antenna unit of the card controlled by the control unit. Visual description of the BTC is given in Figure 5.

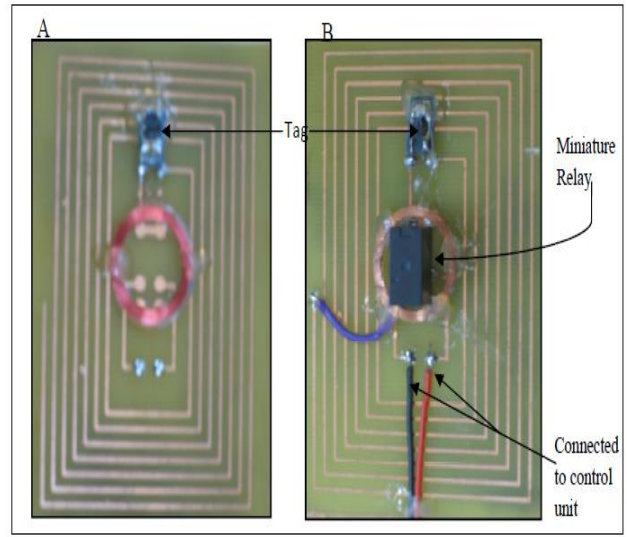

Figure 5: Fabricated Result of Antenna Unit

The control unit was designed using an Atmel AVR-Atmega-8515 microcontroller securely coded using assembly language and AVR studio 4. However, different light emitting diodes (LEDs) were used as indicator on the state of the Card at any given point in operation (see Table 2).

## VI. DISCUSSION

We tested the concept following the procedure stated in Figure 6, and it responded as programmed, practically denying access to unauthorized user.
Furthermore, we subjected BTC to different degree of risk assessment, a process synonymous with fault testing in electronics, or penetration testing in networking environment. In order to evaluate this concept, we demonstrated the following risk assessment processes.
*Tag Manipulation***:** we placed the Card at various angles, proximities and direction to an RFID reader without due authorization from the authentic user. However, there was no interrogation. *Clip joint circumvention***:** We assumed that an attacker could gain access to the internal architecture of the Card (which is practically infeasible). We bridged the clip joint using connecting cables at first, and later using a 5v supply unit.

Table 2: Control Output Indication

| Template label | Atmega-8515 Pin-out | Control Effect | Indication |
|---|---|---|---|
| A and B | PORTB, 6:4 PORTD 7 | Miniature / Green-LED | Authorized user with access permission, access granted |
| C | PORTD, 6 | Yellow-LED | Authorized user without access permission, access denied |
| D | PORTD, 5 | Blue-LED | Unauthorized user, access denied |
| E | PORTD, 4 | Red-LED | Unauthorized user, access denied (and further warning may be indicated |

The former could not initiate the interrogation but the later attempted to trigger the switch trigger (a 5v relay in this case). *Fingerprint manipulation*: We forged an OHP film fingerprint of the residue print on the surface of the scanner. This forges film was then disguised as an authentic user. The evaluation process further proved the security potency of this concept. The fingerprint manipulation could not initiate interrogation due the secured practice exhibited in the requirement for authorization. However, we discovered that unauthorized tag use could be mitigated with this concept. In addition , a securely design process, and a more aligned fabrication process of the clip joint, such attack is practically infeasible or extremely expensive. Other forms of risk associated with the typical RFID Card can thus be successfully mitigated

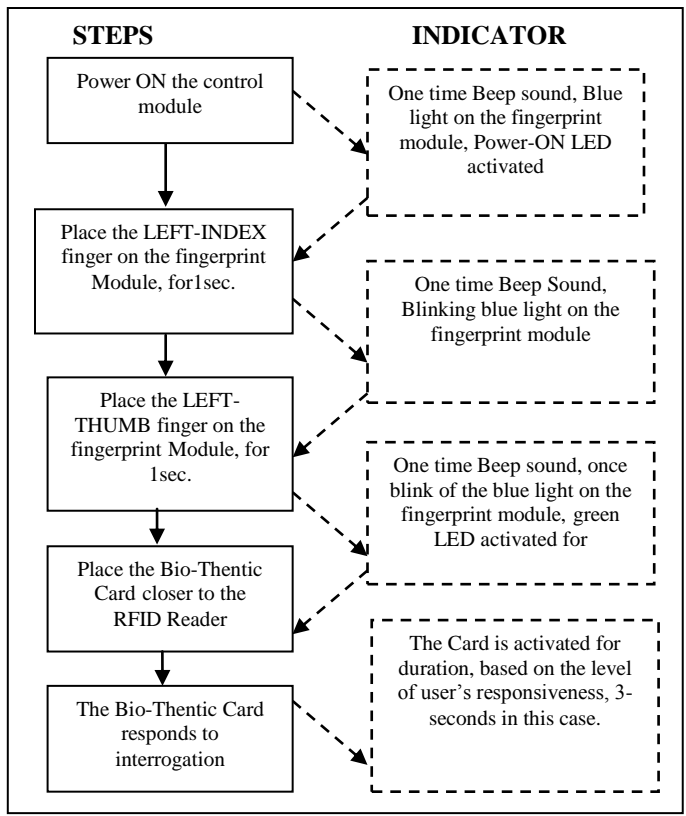

Figure 6a: Testing Procedure for Authentic User

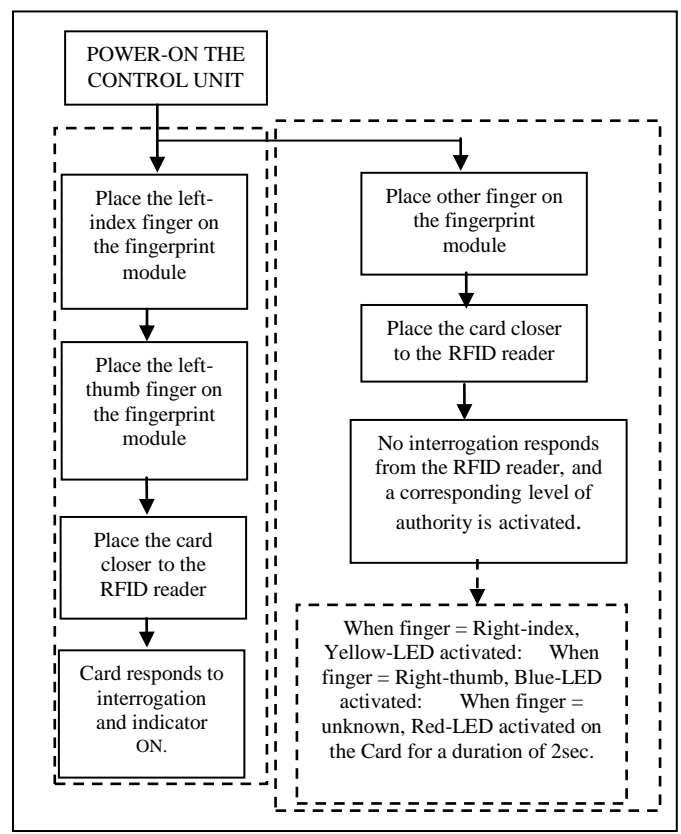

Figure 6b: Generic Testing Procedure

## VII. CONCLUSION

The main contribution of this paper is derived from the research carried out on authentication of an RFID card holder, on the card itself. This is predicated on the fact that the confidentiality of a system that adopts the use of RFID Card is vulnerable to unauthorized use. This paper therefore presents a concept of on-card authentication system as a preventive measure against unauthorized use of RFID Card. An on-card authentication system called Bio-Thentic card was designed, fabricated and evaluated. Furthermore, the Card was subjected to various known attacks, as a risk evaluation measure. The Bio-Thentic card proves to mitigate unauthorized Card use, and consequentially, prevents most known attacks against the RFID Card.